\documentclass[aps,pra,superscriptaddress,twocolumn,10pt,floatfix,showpacs,showkeys]{revtex4-2}

\usepackage{amsmath}
\usepackage{amssymb}
\usepackage{amsthm}
\usepackage{color}
\usepackage{graphicx}
\usepackage{epsfig}
\usepackage{subfigure}
\usepackage{hyperref}
\usepackage[utf8]{inputenc}
\usepackage[T1]{fontenc}
\usepackage{times}
\usepackage{eurosym}
\usepackage[english]{babel}
\usepackage{ulem}
\usepackage{bm}
\usepackage{latexsym}
\usepackage[english]{babel}
\usepackage{natbib}
\usepackage{appendix}
\usepackage{amssymb}

\hypersetup{
    colorlinks=true,linkcolor=blue,citecolor=blue,
    filecolor=blue,urlcolor=blue,breaklinks=true
}

\newcommand{\orcid}[1]{\href{https://orcid.org/#1}{\includegraphics[width=7pt]{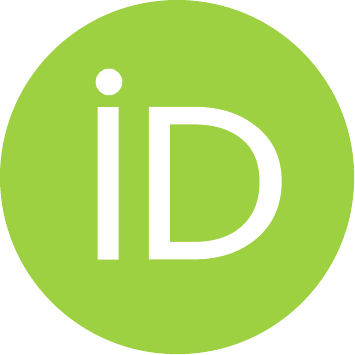}}}

\begin{document}

\title{Mixed states driven by Non-Hermitian Hamiltonians of a nuclear spin ensemble}

\author{D. Cius\orcid{0000-0002-4177-1237}}
\email{danilocius@gmail.com}
\affiliation{
Programa de P\'os-Gradua\c{c}\~{a}o em Ci\^{e}ncias/F\'{i}sica,
Universidade Estadual de Ponta Grossa,
84030-900, Ponta Grossa, Paran\'a, Brazil}
\author{A. Consuelo-Leal\orcid{0000-0003-1141-210X}}
\affiliation{Instituto de F\'{i}sica de S\~{a}o Carlos, Universidade de S\~{a}o Paulo, CP 369, 13560-970, S\~{a}o Carlos, SP, Brasil.}
\author{A. G. Araujo-Ferreira\orcid{0000-0002-6676-384X}}
\affiliation{Instituto de F\'{i}sica de S\~{a}o Carlos, Universidade de S\~{a}o Paulo, CP 369, 13560-970, S\~{a}o Carlos, SP, Brasil.}
\author{R. Auccaise\orcid{0000-0002-9602-6533}}
\email{raestrada@uepg.br}
\affiliation{Departamento de F\'{\i}sica, Universidade Estadual de Ponta Grossa, 84030-900, Ponta Grossa, PR, Brazil.}


\date{\today}

\begin{abstract}
We study the quantum dynamics of a non-interacting spin ensemble under the effect of a   reservoir by applying the framework of the non-Hermitian Hamiltonian operators. Theoretically, the  two-level model describes the quantum spin system and the Bloch vector to establish the dynamical evolution. Experimentally, phosphorous ($^{31}$P) nuclei with spin $I=1/2$  are used to represent the two-level system and the magnetization evolution is measured and used to compare with the theoretical prediction. At room temperature, the composite dynamics of the radio-frequency pulse plus  field inhomogeneities (or unknown longitudinal fluctuations) along the $z$-axis   transform the initial quantum state and drives it into a mixed state at the end of the dynamics. The experimental setup shows a higher  accuracy when compared with the theoretical prediction (>98\%), ensuring the relevance and effectiveness of the non-Hermitian theory at a high-temperature regime.
\end{abstract}

\maketitle

\section{Introduction} \label{sec-int}

Since the foundations of modern quantum theory, the Hermitian property of any operator that yields an observable quantity, which ensures  the compromise between its theoretical physical meaning and its measurability, is deeply related to the operator's real eigenvalues. On the other hand,  the non-Hermitian counterpart has always taken a secondary role. Nevertheless, the non-Hermitian features of some Hamiltonian operators have been receiving more attention and driving a discussion on quantum harmonic oscillators with an extra polynomial imaginary potential energy, where theoretical arguments of $\mathcal{PT}$-symmetry emerge to explain its real spectrum \cite{bender1998}. These results have triggered a series of analogous studies exploring the $\mathcal{PT}$-symmetry as new metric operators definitions in the context of pseudo-hermiticity \cite{mostafazadeh2002A,mostafazadeh2002B,mostafazadeh2002C}, and extending  to different contexts as optical lattices \cite{makris2008}, waveguides \cite{klaiman2008}, driven XXY spin 1/2 chain \cite{prosen2012}, and performing applications like a quantum simulation of the fast evolution of a $\mathcal{PT}$-symmetric Hamiltonian on a two qubit NMR system \cite{zheng2013}. Furthermore, another work used a microwave billiard to analyze $\mathcal{PT}$ symmetry and spontaneous symmetry breaking theoretically and experimentally \cite{bittner2012}. More recently, $\mathcal{PT}$-symmetry breaking was implemented with a single nitrogen-vacancy center in diamond \cite{wu2019}, in a superconducting quantum interference device (SQUID) \cite{naghiloo2019}, hybrid experimental setups as semiconductor devices (InGaAsP platform) and optical pumping to generate topological light-transport channels \cite{zhao2019}. In parallel, an intriguing result raises from the theoretical investigations on time-dependent non-Hermitian Hamiltonians \cite{fring2016a}, as in Refs. \cite{deponte2019,dourado2020,dourado2021a}, where the authors shed light on the possibility of achieving an infinite degree of squeezing into a finite time interval by defining a suitable time-dependent metric. 

Following similar reasoning  as detailed in the previous examples, the analysis of an open quantum system dynamics can be developed considering that the Hamiltonian of the system is non-Hermitian. If this condition is chosen, then some implications arise and must be analyzed in order to preserve quantum principles. There are arguments in studies with atoms/molecules trapped on a solid surface  \cite{moiseyev1998}, photons on cavities \cite{plenio1998}, the mean-field dynamics of a non-Hermitian Bose-Hubbard dimer \cite{graefe2008} that supports this quantum approach. Furthermore, the most explored application of this approach is the analysis performed on the quantum speed limit for non-unitary evolution \cite{taddei2013,deffner2013,del-campo2013,pires2016}, the evolution speed of mixed quantum states \cite{carlini2008,brody2012,garcia-pintos2019}, establishing evolution speed bounds \cite{del-campo2013}, and the exactly solvable damped Jaynes-Cummings model for a two-level system interacting with a bosonic quantum reservoir at zero temperature, considered to describe the quantum speed limit bounds \cite{mirkin2016}.

In many of those applications, the non-Hermitian property of the Hamiltonian operator has been successfully explored to describe the two-level system interacting with a reservoir \cite{mirkin2016}. One of the main common characteristics of those previous theoretical and experimental setups is its low-temperature regime. It implies that, almost always, the quantum system evolves to the ground state and the maximum purity value is achieved. On the other hand, quantum systems  at the high-temperature regime are not frequently analyzed because quantum signatures are drastically diminished. Nevertheless, even with these limiting weak quantum features, the nuclear spin dynamics monitored by the nuclear magnetic resonance preserves its quantumness. In this sense, the dynamics of the  spin system with $I=1/2$, under a time-dependent transversal weak magnetic field and its relaxation behavior, has been studied in different approaches applying the Redfield equation \cite{jones1966,jacquinot1973,smith1992,bull1992,palke1997} and the Bloch equations \cite{viola2000}. In light of those standard mathematical approaches, the application of non-Hermitian Hamiltonians emerges as another alternative physical model to describe decoherence and relaxation dynamics.
 
From the arguments introduced in the previous paragraphs, this study explores the non-Hermitian formalism and its extension to a nuclear spin system in the NMR technique. In order to achieve this task, this manuscript is organized as follows. First, in Section \ref{sec-unst} we introduce the master equation framework and its extension to the non-Hermitian regime in the framework of the Bloch vector. Then, in Section \ref{sec-exp}, we explain the experimental setup: the spectrometer, the samples, the Hamiltonian, and the density matrix of the nuclear spin system. Finally, in Sections \ref{sec-dis} and \ref{sec-conc} we present our main discussions and conclusions, respectively.

\section{Spin Dynamics on non-Hermitian approach} \label{sec-unst}

The primary purpose of our study is to compare experimentally and theoretically the magnetization dynamics of an ensemble of spin-half nuclei submitted to a weak magnetic field $\mathbf{B}_{y}$ along the minus $y$ axis. Ideally, any spin system will precess around the $y$ axis under the unitary transformation that characterize the described setup. Moreover, it is well known that any components of the nuclear magnetization is the sum of innumerable small contributions from the individual spins. Therefore, to obtain the net magnetization, we can apply the successful density matrix method describing the quantum state of the entire ensemble, without referring to the individual spin states \cite{levitt2008}.

For a sample of spin-half nuclei, given the nuclear symmetry properties and appropriate concentration of nuclei in the sample, it can be assumed that the $N$ spins in the ensemble do not interact amongst themselves, even at room temperature. Thus, the Redfield approach seems more proper to precisely describe the dynamical properties of a nuclear spin ensemble \cite{nicholas2010}, which is mathematically described through master equations \cite{redfield1957}. However, non-unitary processes occur during the NMR experiments, and a rigorous description of spin dynamics may not be an easy task from the mathematical point of view due to the complexity of the reservoir.

Therefore, instead of the Redfield approach, we consider a simple model based on the effective non-Hermitian Hamiltonian to obtain a phenomenological description of the time evolution of the magnetization components. Certainly, the direct application of the non-Hermitian Liouville-von Neumann equation
\begin{equation*}
i\hbar \partial_{t}\hat{\rho}=\hat{H}_{\text{eff}}\hat{\rho}-\hat{\rho}\hat{H}_{\text{eff}}^{\dagger } \, ,
\end{equation*}%
might not make sense and cause some confusion when the trace of the density matrix is not preserved in time. This entails probability losses and puts the statistical meaning in an obscure scenario according to the standard quantum formalism \cite{holevo2001}. However,  D. C. Brody and E. Graefe \cite{brody2012} and A. Sergi and K. G. Zloshchastiev \cite{sergi2013} considered the density matrix of a two-level system and  studied the non-Hermitian time evolution in order to mimic the coupling with a dissipative environment. They proposed the following non-linear Liouville-von Neumann equation
\begin{equation}
i\hbar \partial_{t} \hat{\varrho}= [ \hat{H}_{0},\hat{\varrho} ] - i \{ \hat{\Gamma}_{0},\hat{\varrho} \}
+ 2i\,\mathrm{\mathbf{Tr}}[\, \hat{\Gamma}_{0}\hat{\varrho} \,] \hat{\varrho} \,
\label{Sergi}
\end{equation}
by introducing a normalized density operator $\hat{\varrho} = \hat{\rho}/\mathrm{\mathbf{Tr}}\hat{\rho}$, and considering the  effective non-Hermitian Hamiltonian in the form $\hat{H}_{\text{eff}}=\hat{H}_{0}-i\hat{\Gamma}_{0}$, with both $\hat{H}_{0}$ and $\hat{\Gamma}_{0}$ being Hermitian operators. The non-linear term $2i\,\mathrm{\mathbf{Tr}}[\, \hat{\Gamma}_{0}\hat{\varrho}\, ] \hat{\varrho}$ arises from the time derivative of $\mathrm{\mathbf{Tr}}\hat{\rho}$, and it is accountable for the trace-preserving time-evolution of $\hat{\varrho}$. Eq. (\ref{Sergi}) has been broadly applied and deeply discussed in different theoretical scenarios such as symmetry breaking \cite{brody2012}, criticality in $\mathcal{PT}$-symmetric systems \cite{kawabata2017}, quantum speed limits \cite{taddei2013,pires2016} and others \cite{brody2013,sergi2015}.

From some of those discussions, Eq. (\ref{Sergi}) can be rewritten as follows
\begin{equation}
i\hbar \partial_{t}\hat{\varrho}=[ \hat{H}_{0}, \hat{\varrho}]-i \{ \hat{\Gamma}_{\hat{\varrho}}^{0}(t),\hat{\varrho} \}, \label{eqas}
\end{equation}
where we define the operator $\hat{\Gamma}_{\hat{\varrho}}^{0}(t)=\hat{\Gamma}_{0} - \mathrm{\mathbf{Tr}}[\, \hat{\Gamma}_{0}\hat{\varrho}\,] \hat{\boldsymbol{1}}$. Therefore, this master equation points out an effective non-Hermitian Hamiltonian
\begin{equation}
\hat{H}_{\text{eff}}(t)=\hat{H}_{0}-i\hat{\Gamma}_{0}(\hat{\varrho})\,.
\label{Heff}
\end{equation}
Notice that the term $\mathrm{\mathbf{Tr}}[\, \hat{\Gamma}_{0}\hat{\varrho}\,]$ introduces a state-dependence to the effective Hamiltonian  which reflects a kind of feedback over the dynamics as happens in a mean-field approximation to superradiant emission  \cite{mizrahi1990,*mizrahi1993,dourado2021b}. The Gisin equation is achieved from Eq. \eqref{eqas} by considering $\hat{\Gamma}_{0}=\hat{H}_{0}$ as a candidate to describe dissipative quantum dynamics \cite{gisin1981}. Furthermore, R. Wieser \cite{wieser14,*wieser14e} showed that the classical Landau-Lifshitz equation can be derived from the quantum mechanical description provided by  Eq. \eqref{eqas}, assuming certain considerations.

The Hermitian Hamiltonian $\hat{H}_{0}$ of a two-level quantum system  can be written as
\begin{equation}
\hat{H}_{0}=\hbar\omega_{x}\hat{\mathbf{I}}_{x} + \hbar\omega_{y}\hat{\mathbf{I}}_{y} + \hbar\omega_{z}\hat{\mathbf{I}}_{z}\,,
\label{H0}
\end{equation}
where $\omega_{k}$ represents the $k$-th angular frequency component, and the basis operators are given by $\hat{\mathbf{I}}_{0}=\hat{\boldsymbol{1}}/2$ and $\hat{\mathbf{I}}_{k}=\hat{\boldsymbol{\sigma}}_{k}/2$  where  $\hat{\boldsymbol{\sigma}}_{k}$ represents the $k$-th Pauli matrix $(k=x,y,z)$. In addition, we generalize the Sergi's approach by including the  \textit{ ad hoc} time-dependent operator $\hat{\Gamma}_{0}\rightarrow\hat{\Gamma}(t)$:
\begin{equation}
i\hbar \partial_{t}\hat{\varrho}=[ \hat{H}_{0}, \hat{\varrho}]-i \{ \hat{\Gamma}_{\hat{\varrho}}(t),\hat{\varrho} \}, \label{eqasg}
\end{equation}
with $\hat{\Gamma}_{\hat{\varrho}}(t) =\hat{\Gamma}(t)- 2\mathrm{\mathbf{Tr}}[\, \hat{\Gamma}\hat{\varrho}\,]\hat{\mathbf{I}}_{0}$ in which we assume the operator $\hat{\Gamma}(t)$ in the general form
\begin{align}
\hat{\Gamma}(t)= \hbar\lambda_{0}(t) \hat{\mathbf{I}}_{0}  + \hbar \lambda_{x}(t)  \hat{\mathbf{I}}_{x} + \hbar \lambda_{y}(t) \hat{\mathbf{I}}_{y} + \hbar \lambda_{z}(t) \hat{\mathbf{I}}_{z} \, ,
\label{Ggen}
\end{align}
where the $\lambda_{i}$'s are real functions to be determined and they are associated with the reservoir in a phenomenological sense. Looking at the operator $\hat{\Gamma}(\hat{\varrho})$, we note that the first term in Eq. (\ref{Ggen}) does not actually contribute to the dynamics and it can be neglected.

The general quantum state of a two-level system is given by
\begin{equation}
\hat{\varrho}(t)=\hat{\mathbf{I}}_{0}+ \mathrm{r}_{x}(t)\hat{\mathbf{I}}_{x} + \mathrm{r}_{y}(t)\hat{\mathbf{I}}_{y} + \mathrm{r}_{z}(t)\hat{\mathbf{I}}_{z}, \label{rhogen}
\end{equation}
with the time-dependent coefficients
\begin{equation}
\mathrm{r}_{k}(t) = 2\mathrm{\mathbf{Tr}}[\, \hat{\mathbf{I}}_{k}\hat{\varrho}(t)\,], \label{BlochVec}
\end{equation}
being the $k$-th component of the Bloch vector ($k=x,y,z$). We now apply the quantum state (\ref{rhogen}) and the operator (\ref{Ggen}) in the  non-linear non-Hermitian Liouville-von Neumann equation (\ref{eqas}), and obtain a set of differential equations to be satisfied by $\mathrm{r}_{k}(t)$:
\begin{widetext}
\begin{subequations}
\label{Diffequation}
\begin{align}
\partial_{t}\mathrm{r}_{x}(t)& = \mathrm{r}_{x}(t)\left[\lambda_{x}(t)\mathrm{r}_{x}(t)+\lambda_{y}(t)\mathrm{r}_{y}(t)+\lambda_{z}(t)\mathrm{r}_{z}(t)\right] +\omega_{y}\mathrm{r}_{z}(t) -\omega_{z}\mathrm{r}_{y}(t)-\lambda_{x}(t), \\
\partial_{t}\mathrm{r}_{y}(t)& = \mathrm{r}_{y}(t)\left[\lambda_{x}(t)\mathrm{r}_{x}(t)+\lambda_{y}(t)\mathrm{r}_{y}(t)+\lambda_{z}(t)\mathrm{r}_{z}(t)\right] +\omega_{z}\mathrm{r}_{x}(t) -\omega_{x}\mathrm{r}_{z}(t)-\lambda_{y}(t), \\
\partial_{t}\mathrm{r}_{z}(t)& = \mathrm{r}_{z}(t)\left[\lambda_{x}(t)\mathrm{r}_{x}(t)+\lambda_{y}(t)\mathrm{r}_{y}(t)+\lambda_{z}(t)\mathrm{r}_{z}(t)\right] +\omega_{x}\mathrm{r}_{y}(t) -\omega_{y}\mathrm{r}_{x}(t)-\lambda_{z}(t).
\end{align}
\end{subequations}
\end{widetext}
In order to describe the depolarization effects, decreasing the amplitude of the magnetization components, we introduce a decay as a real positive function $f(t)$. It determines the decreasing rate in the magnitude of the components of the Bloch vector along the time. Thus, we infer the solutions having the form
\begin{align}
\mathrm{r}_{k}(t) = f(t) \mathrm{r}_{k}^{0}(t), \qquad k=x,y,z \, ,  \label{DampingBloch} 
\end{align}
where we are considering $f(0)=1$, whereas $\mathrm{r}_{k}^{0}(t)$  is the $k$-th component of the Bloch vector in the absence of any loss processes, and satisfying the condition
\begin{equation}
\left[\mathrm{r}_{x}^{0}(t)\right]^{2} + \left[\mathrm{r}_{y}^{0}(t)\right]^{2} + \left[\mathrm{r}_{z}^{0}(t)\right]^{2}= \left[\mathrm{r}^{0}(0)\right]^{2} , \label{NormalizationMagnetization}
\end{equation}
with $\mathrm{r}^{0}(0)$ being a constant fixed at initial time corresponding to the Bloch sphere radius at $t=0$. For initially pure states, the radius of the Bloch sphere is equal to the unit; otherwise, for mixed states, it assumes a positive value smaller than the unit.  
Furthermore, we suitably choose the $\lambda$'s functions assuming that they are determined from the coherent magnetization components $ \mathrm{r}_{x,y,z}^{0}(t)$ according to
\begin{align}
\lambda_{k}(t)& = g(t) \mathrm{r}_{k}^{0}(t), \qquad k=x,y,z \,, \label{Dampingfunctions}
\end{align}
we assume the interaction is turned on at $t = 0$ and no exchanges exists between the system and the reservoir at this initial time. Thus, we have $g(0)=0$. At later times, correlations between the system and the reservoir will arise due to the coupling, and in this case $g(t)$ assumes non-null values. The components $\mathrm{r}_{x,y,z}^{0}(t)$ are bounded and behave harmonically in time. In this approach, the function $g(t)$ works modulating these components, and the assumption proposed in Eq. (\ref{Dampingfunctions}) suggests that the depolarization effects in each component depend on the Bloch vector components and decrease in time. Now we consider the solutions (\ref{DampingBloch}) and the functions (\ref{Dampingfunctions}) in the differential equations (\ref{Diffequation}), and taking into account the condition (\ref{NormalizationMagnetization}) which is satisfied by $\mathrm{r}_{x,y,z}^{0}(t)$. Straightforwardly, Eqs. (\ref{Diffequation}) become
\begin{subequations}
\label{coherent}
\begin{align}
\partial_{t}\mathrm{r}_{x}^{0}(t)& = \omega_{y}\mathrm{r}_{z}^{0}(t) - \omega_{z}\mathrm{r}_{y}^{0}(t), \\
\partial_{t}\mathrm{r}_{y}^{0}(t)& = \omega_{z}\mathrm{r}_{x}^{0}(t) - \omega_{x}\mathrm{r}_{z}^{0}(t), \\
\partial_{t}\mathrm{r}_{z}^{0}(t)& = \omega_{x}\mathrm{r}_{y}^{0}(t) - \omega_{y}\mathrm{r}_{x}^{0}(t),
\end{align}	
\end{subequations}
by imposing the constraint
\begin{equation}
g(t)=\frac{\partial_{t}f(t)}{\left[\mathrm{r}^{0}(0)f(t)\right]^{2}-1}, \quad t>0, \label{geq}
\end{equation}
in order to describe the coherent dynamics for the $\mathrm{r}_{k}^{0}(t)$ components. By considering the system initially prepared in the pure state corresponding to $\mathrm{r}_{z}^{0}(0)=1$ and $\mathrm{r}_{x,y}^{0}(0)=0$, the solutions of Eq. (\ref{coherent}) assume the form
\begin{subequations}
\label{BlochComp}
\begin{align}
\mathrm{r}_{x}^{0}(t) &=\frac{\omega_{x}\omega_{z}}{\Omega^{2}}\left[1-\cos{\left(\Omega t\right)} \right] 
                       + \frac{\omega_{y}}{\Omega}\sin{\left(\Omega t\right)},  \\
\mathrm{r}_{y}^{0}(t) &=\frac{\omega_{y}\omega_{z}}{\Omega^{2}}\left[1-\cos{\left(\Omega t\right)} \right] 
                       - \frac{\omega_{x}}{\Omega}\sin{\left(\Omega t\right)}, \\
\mathrm{r}_{z}^{0}(t) &=\frac{\omega_{z}^{2}}{\Omega^{2}}\left[1-\cos{\left(\Omega t\right)} \right] + \cos{\left(\Omega t\right)},
\end{align}
\end{subequations}
with the effective angular frequency $\Omega^{2} = \omega_{x}^{2} + \omega_{y}^{2} + \omega_{z}^{2}$. 

Since the depolarization effect is described by the operator (\ref{Ggen}) in the non-Hermitian approach, we also suppose that it effectively acts in a finite time interval due to the non-linear feature of the relaxation process. This means that the system does not achieve a maximally mixed state, and to precisely describe the ensemble dynamics, we introduce the following \textit{Ansatz} for the decay function $f(t)$:
\begin{equation}
f(t) = e^{-\delta t} + \nu\left(1-e^{-\mu t}\right),
\end{equation}
where $\mu$ and $\delta$ are the decay parameters, and we call $\nu$ the Bloch sphere radius for a long time interval such that $\nu \ll 1$. Consequently, the function $g(t)$ in (\ref{geq}) becomes
\begin{equation}
g(t)=\frac{ \delta e^{-\delta t} -\nu\mu e^{-\nu t}}{1-\left[e^{-\delta t} + \nu\left(1-e^{-\mu t}\right)\right]^{2}} \, . 
\label{g}
\end{equation}
For long time intervals, such that $\mu t,\delta t\gg 1$, the $k$-th component of the Bloch vector behaves according to
\begin{align}
\mathrm{r}_{k}(t) & \approx \nu\mathrm{r}_{k}^{0}(t) ,
\label{BlochComp-longtimes}
\end{align}
which corresponds to a mixed state regime where the Bloch vector oscillates with a very small constant amplitude. To further illustrate this behavior, we apply the purity of the quantum state defined as $\mathsf{P}(t)=\mathrm{\mathbf{Tr}}[\,\hat{\varrho}^{2}(t)\,]$, which can be rewritten as follows
\begin{equation}
\mathsf{P}(t) =  \frac{1}{2}\left[ 1 + \boldsymbol{\mathrm{r}}^{2}(t) \right], \label{purity}
\end{equation}
where $\boldsymbol{\mathrm{r}}^{2}(t)=\mathrm{r}_{x}^{2}(t) + \mathrm{r}_{y}^{2}(t) + \mathrm{r}_{z}^{2}(t)$. Note the purity is bounded, and it is defined in the interval $1/2\leqslant \mathsf{P}(t)\leqslant 1$, where the upper and lower bound corresponds to the pure and the maximally mixed states, respectively. Substituting Eq. (\ref{DampingBloch}) into Eq. (\ref{purity}), we have
\begin{equation}
\mathsf{P}(t) =   \frac{1}{2} + \frac{1}{2}\left[e^{-\delta t} + \nu\left(1-e^{-\mu t}\right) \right]^{2}, \label{purityf}
\end{equation}
which for a long time regime, it approximates to
\begin{equation}
\mathsf{P}(t) \approx   \frac{1}{2} + \frac{\nu^{2}}{2},   \label{purityf-longtimes}
\end{equation}
and from this expression it is understood that the system evolves to a mixed state with very low purity  $\mathsf{P}(t)\gtrsim 1/2$.

We find that the non-unitary dynamics described by the effective non-Hermitian Hamiltonian $\hat{H}_{\text{eff}}(t)=\hat{H}_{0}-i\hat{\Gamma}_{\hat{\varrho}}(t)$, has the depolarization effects described in the framework of non-Hermitian operators by the non-linear operator
\begin{equation}
\hat{\Gamma}_{\hat{\varrho}}(t)=\hbar g(t)\left[ \mathrm{r}_{x}^{0}(t) \hat{\mathbf{I}}_{x} + \mathrm{r}_{y}^{0}(t) \hat{\mathbf{I}}_{y}+\mathrm{r}_{z}^{0}(t)\hat{\mathbf{I}}_{z}\right],
\end{equation}
with $g(t)$ given by Eq. (\ref{g}) for $t>0$. The time-dependent quantum state dynamic generated by the theory of this study is  similar to the extended model discussed in Ref. \cite{brody2012}, where the system evolves from an initially pure state to a mixed state due to noise, but here the state remains oscillating around the fixed point $\hat{\varrho}=\hat{\mathbf{I}}_{0}$ with an amplitude $\nu$, i.e., for long times the quantum state evolves to  $\hat{\varrho}=\hat{\mathbf{I}}_{0} + \nu [ \mathrm{r}_{x}^{0}(t)\hat{\mathbf{I}}_{x}+\mathrm{r}_{y}^{0}(t)\hat{\mathbf{I}}_{y}+\mathrm{r}_{z}^{0}(t)\hat{\mathbf{I}}_{z}]$. In the next section, we apply this approach to describe two experiments in an NMR setup.

\section{Experimental setup : Two-level spin ensemble} \label{sec-exp}

\begin{figure}[!ht]
  \begin{center}
    $\begin{array}{c@{\hspace{0.07in}}c}
     \multicolumn{1}{c}{\mbox{\bf }} &
     \multicolumn{1}{c}{\mbox{\bf }} \\ [-0.53cm]
     \epsfxsize=1.600in
     \epsffile{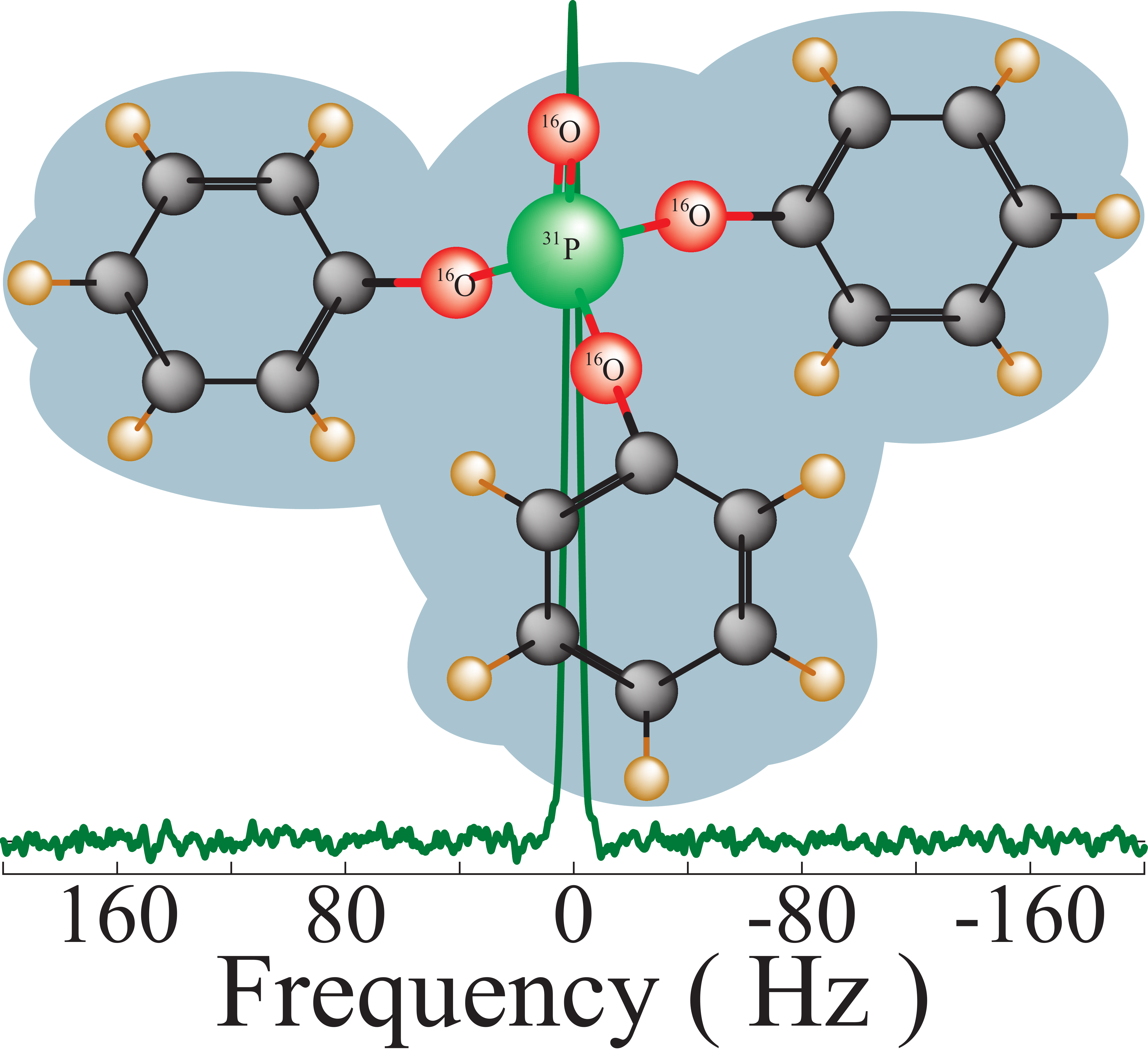} &
     \epsfxsize=1.600in
     \epsffile{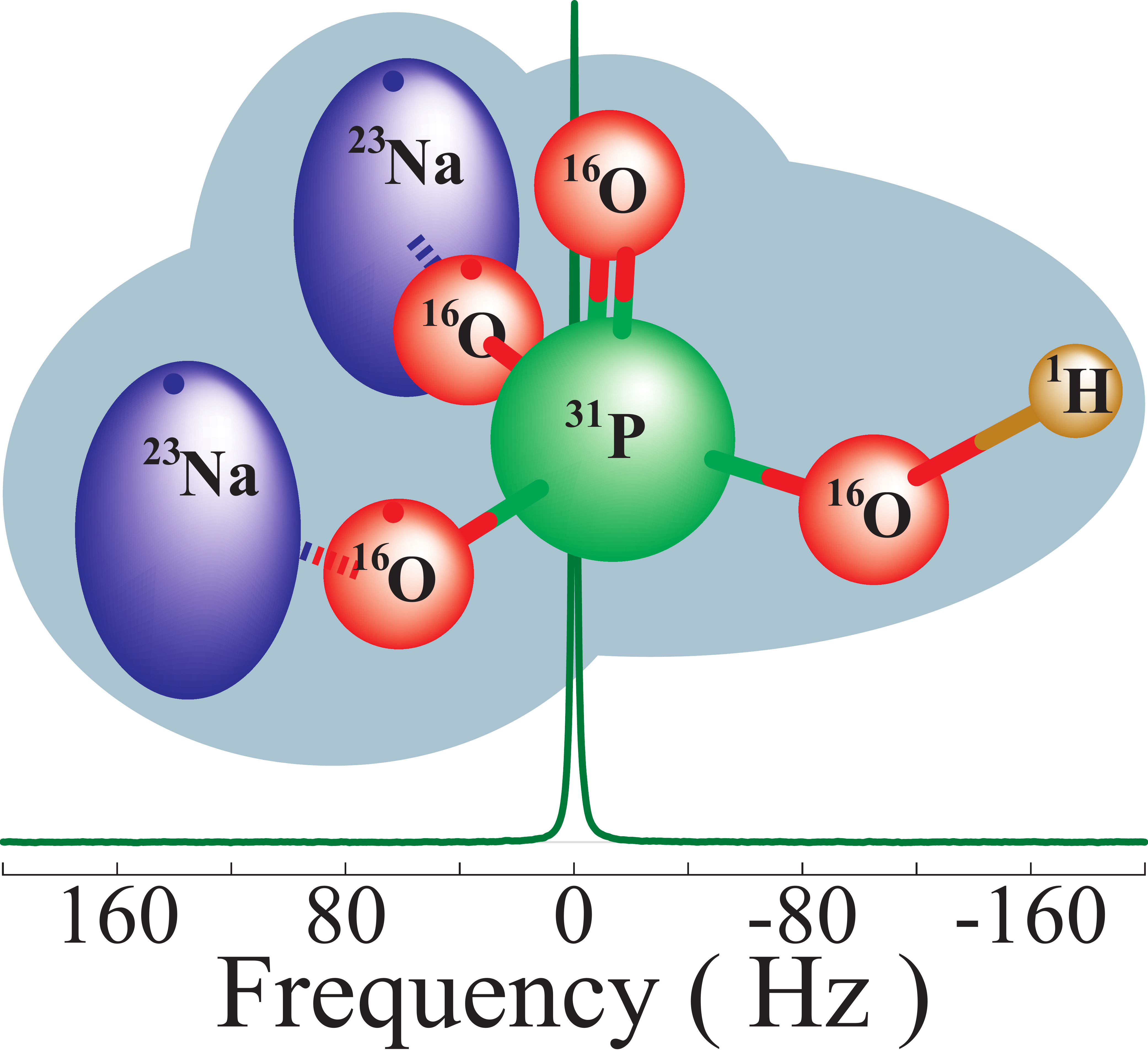} \\ [0.0cm]
     \mbox{ \textbf{(a)} } & \mbox{ \textbf{(b)} }
     \end{array}$
   $\begin{array}{c}
     \multicolumn{1}{c}{\mbox{\bf }} \\ [-0.30cm]
     \epsfxsize=3.3000in
     \epsffile{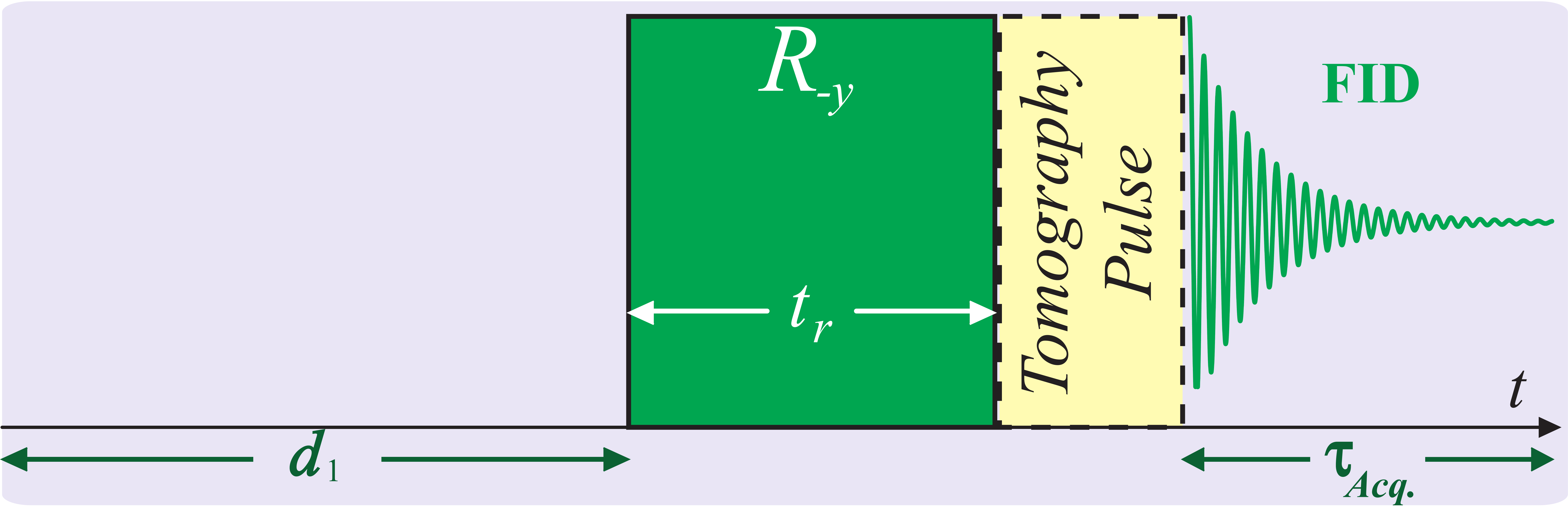} \\ [0.0cm]
     \mbox{ \textbf{(c)} } 
     \end{array}$
  \end{center}
\caption{(Color online) Drawings of the molecular structures of the samples used in the present study (a) Tri-phenyl Phosphate (TPP) C$_{18}$H$_{15}$PO$_{4}$ and the  spectrum of $^{31}$P nuclei are sketched. (b)  Di-sodium Phosphate (DSP) Na$_{2}$HPO$_{4}$  and the spectrum of $^{31}$P nuclei are sketched. The green spheres represent  $^{31}$P nuclei, red spheres represent $^{16}$O nuclei, black spheres represent $^{12}$C nuclei, brown spheres represent  $^{1}$H nuclei and blue ellipsoids represent $^{23}$Na nuclei. (c) The standard pre-saturation pulse sequence was modified in order to implement and to monitor the dynamics of the $^{31}$P nuclei of the sample under a non-Hermitian Hamiltonian. The pulse sequence is divided into four stages, it starts with a recycle delay $d_{1}$ in order to achieve the quantum steady state of nuclei in the sample. The second stage is devoted to implement the Rabi regime using the transformation rotation $\hat{R}_{-y}\left(t_{r}\right)$. The third stage corresponds with the implementation of the tomography procedure denoted by the dashed square, and at the fourth stage is performed the  read out of the free induction decay (FID). }
\label{fig:Moleculas}
\end{figure}

A single two-level quantum model is implemented experimentally on an ensemble of nuclear spin systems by the NMR technique. In this sense, the $^{31}$P nuclei of the tri-phenyl phosphate (TPP) C$_{18}$H$_{15}$PO$_{4}$ molecule, see Fig. \ref{fig:Moleculas}(a), and di-sodium phosphate (DSP) Na$_{2}$HPO$_{4}$  molecule, see Fig. \ref{fig:Moleculas}(b),  are used to represent the quantum model. The tri-phenyl phosphate sample was prepared at  stoichiometric proportions of 0.0485 M tri-phenyl phosphate and dissolved in acetone-d$_{6}$. The di-sodium phosphate sample was prepared at stoichiometry proportions of 12.609 M di-sodium phosphate and dissolved in deuterated water. Each sample solution (solute and solvent) was placed in a 5 mm NMR tube. The sample is placed in a homogeneous strong static magnetic field of $B_{0}=9.39$ T and oriented along the $z$-axis of a spatial coordinate frame.

The NMR spectrometer used in this implementation is a 400 MHz - Ascend III Bruker configured for liquid samples. A multinuclear 5 mm double resonance broadband liquid probe-head with Variable temperature facility, at liquid configuration supplied by a two-channel probe-head encoded by (H/F)X: one channel to detect/excite $^{1}$H or $^{19}$F  nuclei signals and the other one to detect/excite from $^{31}$P until $^{15}$N nuclei signals. The $^{31}$P control set-up for the sample of tri-phenyl phosphate run at the radio-frequency $\omega_{\text{rf}}=2 \pi \left( 161.973\, \text{MHz} \right)   $, $\pi / 2 $-pulse time is 11.80 $\mu$s or equivalently $\omega_{1} = 2 \pi ( 21\,186 \,\text{Hz} ) $, recycle delay is $d_{1}=80$s, acquisition time is $\tau_{\text{Acq.}}=0.8$s. Also, the transversal relaxation time ($T_{2}$) and the longitudinal relaxation time   ($T_{1}$) were experimentally measured and found to be  $1.0\ \text{s}$ and $8.1\ $s, respectively. The $^{31}$P control setup for the sample of di-sodium phosphate run at the radio-frequency $\omega_{\text{rf}}=2 \pi \left( 161.976\, \text{MHz} \right)   $, $\pi / 2 $-pulse time is 13.40 $\mu$s or equivalently $\omega_{1} = 2 \pi ( 18\,657 \, \text{Hz} ) $, recycle delay is $d_{1}=80$s, acquisition time is $\tau_{\text{Acq.}}=2.0$s.  Also, the transversal relaxation time ($T_{2}$) and the longitudinal relaxation time   ($T_{1}$) were experimentally measured and found to be $0.51\ \text{s}$ and $12.3\ $s, respectively. Using these experimental parameters, the $^{31}$P signals of the TPP and the DSP samples  were measured with only one scan and the spectra are shown in Fig. \ref{fig:Moleculas}a and Fig. \ref{fig:Moleculas}b, respectively. Significant and qualitative differences between both spectra point out that  the  $^{31}$P spectrum of the TPP sample is noisier when compared with  the  $^{31}$P spectrum of the DSP sample. This characteristic matches with the stoichiometric relation between sample-solvent for TPP and DSP samples. Therefore, from now onwards, the experimental implementations were performed using two scans for the   $^{31}$P signal of the TPP sample and using one scan for the   $^{31}$P of the DSP sample.

Describing this setup in the laboratory frame, the magnetic moment of the nuclear spin system interacts with a homogeneous magnetic field  $B_{0}$ aligned along the $z$-axis establishing the Zeeman energy, or the secular Hamiltonian, as the first energy contribution. This Hamiltonian is denoted by $\hat{H}_{\text{s}}=-\hbar\gamma B_{0}\hat{\mathbf{I}}_{z}=-\hbar\omega_{\text{L}}\hat{\mathbf{I}}_{z}$ where $\gamma$ is the gyromagnetic ratio, $\omega_{\text{L}}$ is the Larmor frequency, and $\hat{\mathbf{I}}_{k}$ the nuclear spin operators. Another time-dependent weak magnetic field  $B_{1}$  parallel to the $xy$-plane interacts with the magnetic moment of the nuclei establishing the radio-frequency energy, the second energy contribution, denoted by $\hat{H}_{\text{rf}} \left( t  \right)  =\hbar \omega_{1} [ \cos\left( \omega_{\text{rf}} t + \phi \right)\hat{\mathbf{I}}_{x} + \sin\left( \omega_{\text{rf}} t + \phi \right)\hat{\mathbf{I}}_{y} ]$, with the radio-frequency strength defined by $ \omega_{1}= \gamma B_{1}$. The total energy of the nuclear spin system is denoted by
\begin{equation}
\hat{H}  \left( t  \right) =   \hat{H}_{\text{s}}   +  \hat{H}_{\text{rf}} \left( t  \right) \text{,}
\end{equation}
such that the Hamiltonian at the rotating frame reads as
\begin{equation}
\hat{H}=-\hbar (\omega_{\text{L}}-\omega_{\text{rf}})\hat{\mathbf{I}}_{z} +  \hbar \omega_{1} ( \cos{\phi}\, \hat{\mathbf{I}}_{x} +  \sin{\phi}\, \hat{\mathbf{I}}_{y}  )  \text{.} \label{expH}
\end{equation}
The generation of the effective Hamiltonian is achieved on resonance  $\omega_{\text{rf}}=\omega_{\text{L}}$ and precessing around an effective magnetic field along the negative $y$ axis with $\phi=3\pi/2$ such that the Hamiltonian of the Eq. (\ref{expH}) might be written as $ \hat{H} =-\hbar\omega_{1}\hat{\mathbf{I}}_{y}$. Thus, the target quantum state is transformed using a rotation operator defined by
\begin{equation}
\hat{R}_{-y}\left(t_{r}\right) = \exp\left[ - \frac{i}{\hbar}  \hat{H} \,  t_{r} \right]  =\exp\left[  i\omega_{1}t_{r}\hat{\mathbf{I}}_{y}\right] \text{.} \label{Utr}
\end{equation}

The most appropriate procedure to experimentally describe any ensemble of nuclear spins is using the density operator. Following principles of statistical mechanics, we initially prepared the system in the thermal equilibrium quantum state $\hat{\varrho}_{\text{eq}}$ at temperature $T$. It is represented by the density operator $\hat{\varrho}_{\text{eq}}$ written as
\begin{equation}
\hat{\varrho}_{\text{eq}}=\frac{e^{\beta \hbar \omega_{\text{L}} \hat{\mathbf{I}}_{z}}}{\mathcal{Z}}= \hat{\mathbf{I}}_{0} + \varepsilon(T)\hat{\mathbf{I}}_{z},
\label{thermalstate}
\end{equation}
where $\mathcal{Z}=2\cosh{(\beta \hbar\omega_{\text{L}}/2)}$ is the canonical partition function, and $\beta^{-1}= k_{\text{B}}T$ ( $k_{\text{B}}$ is the Boltzmann's constant), and  
\begin{equation}
\varepsilon(T)=\tanh{\left[\frac{\hbar\omega_{\text{L}}}{2k_{\text{B}}T}\right]}, \label{PolarizationFactor}
\end{equation}
is the so-called polarization factor. 
Note the equilibrium state in Eq. \eqref{thermalstate} can be rewrite in the suitable form 
\begin{equation}
\hat{\varrho}_{\text{eq}} =  [1 - \varepsilon(T)] \hat{\mathbf{I}}_{0} + \varepsilon(T)\hat{\varrho}(0) , 
\label{pseudopurestate}
\end{equation}
where $\hat{\varrho}(0)$ corresponds to the initially pure state described by Eq. \eqref{rhogen} for $\mathrm{r}_{x,y}(0)=0$ and $\mathrm{r}_{z}(0)=1$. Further, the pure state is achieved only at extremely low temperatures in which $\varepsilon(T)\approx1$. However, a remarkable feature of the NMR technique is the possibility of working at a high-temperature regime by preparing of a pseudo-pure state. In this sense, the thermodynamic stationary quantum state (\ref{thermalstate}) of any NMR spin system at high temperature regime has $\varepsilon(T)$ determined by the first order term of a Taylor's expansion of \eqref{PolarizationFactor} as  \cite{oliveira2007}
\begin{equation}
\varepsilon(T) \approx\frac{ \hbar\omega_{\text{L}}}{2k_{\text{B}}T},
\label{NMR_factor}
\end{equation}
where $T$ means the room temperature at $24^{\circ}$C,  $\omega_{\text{L}}  $ is the Larmor frequency for $^{31}$P nuclei at $B_{0}=9.39$ T.
Furthermore, the term proportional to the identity does not contribute to the measured signal since the NMR observables $\hat{\mathbf{I}}_{k}$'s have null trace. It means that the dynamical behaviour of the pseudo-pure state is the same as that of a pure state, and all measurements and experimental data are proportional to the polarization factor as denoted by Eq. (\ref{NMR_factor}) and at the high-temperature regime is $\varepsilon(T)  \approx  1.304\times10^{-5}$.  Moreover, the $z$-component of the spin angular momentum operator $\hat{\mathbf{I}}_{z}$ represents the so-called deviation density matrix $\Delta\hat{\varrho}_{0}$ \cite{oliveira2007},  which is related with thermal state (\ref{thermalstate}) as
$
\Delta\hat{\varrho}_{0} = \hat{\mathbf{I}}_{z} = (\hat{\varrho}_{\text{eq}} - \hat{\mathbf{I}}_{0})/\varepsilon (T)
$, 
which is traceless. This deviation density matrix can be tomographed and reconstructed using global rotations \cite{teles2007}, and this procedure was explained in Ref. \cite{villamizar2018} and applied on a similar spin system.

\begin{figure*}[!ht]
 \begin{center}
    $\begin{array}{c@{\hspace{0.1in}}c}
     \multicolumn{1}{c}{\mbox{\bf }} &
     \multicolumn{1}{c}{\mbox{\bf }} \\ [-0.53cm]
     \epsfxsize=3.3500in
     \epsffile{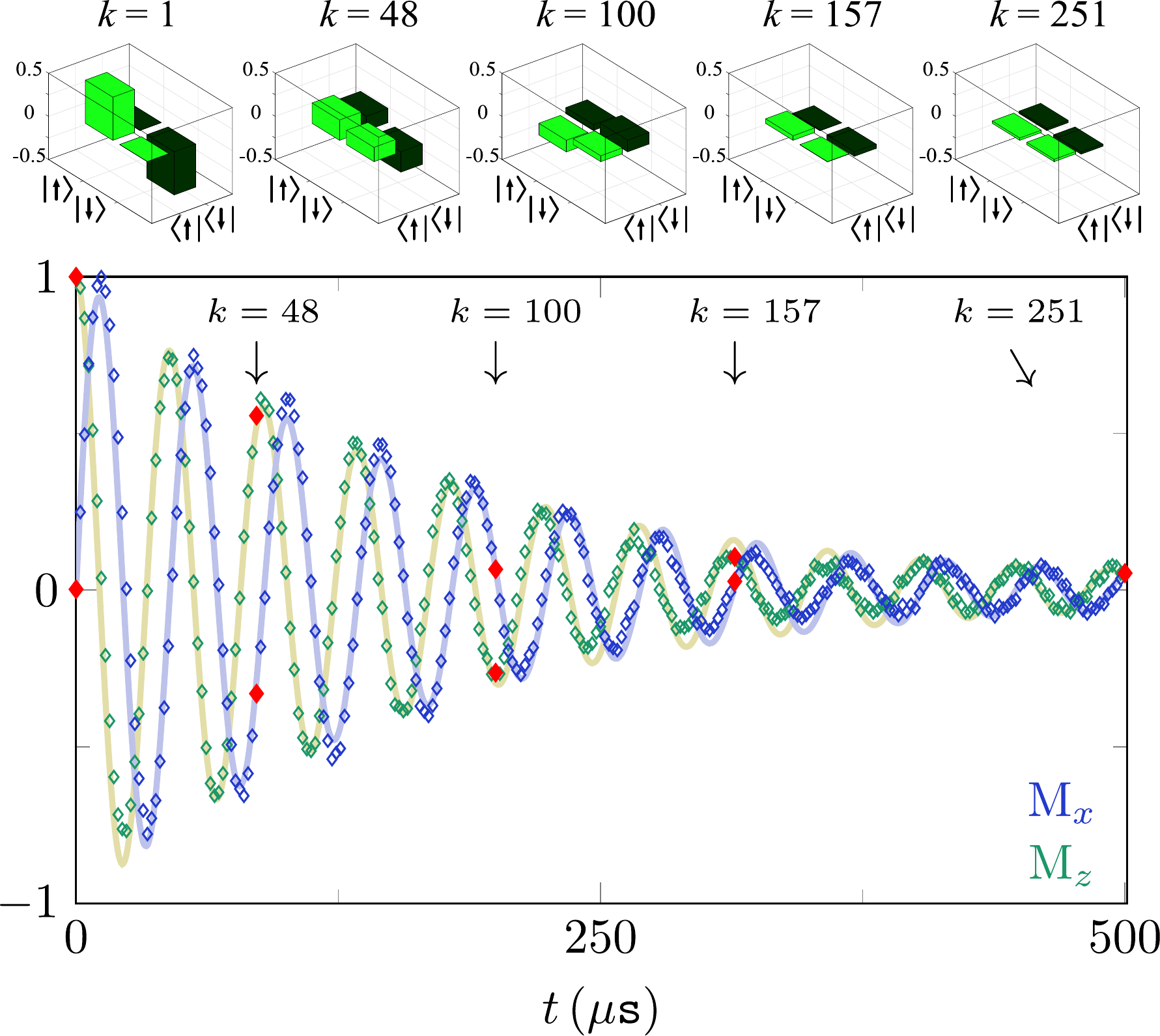} &
     \epsfxsize=3.3500in
     \epsffile{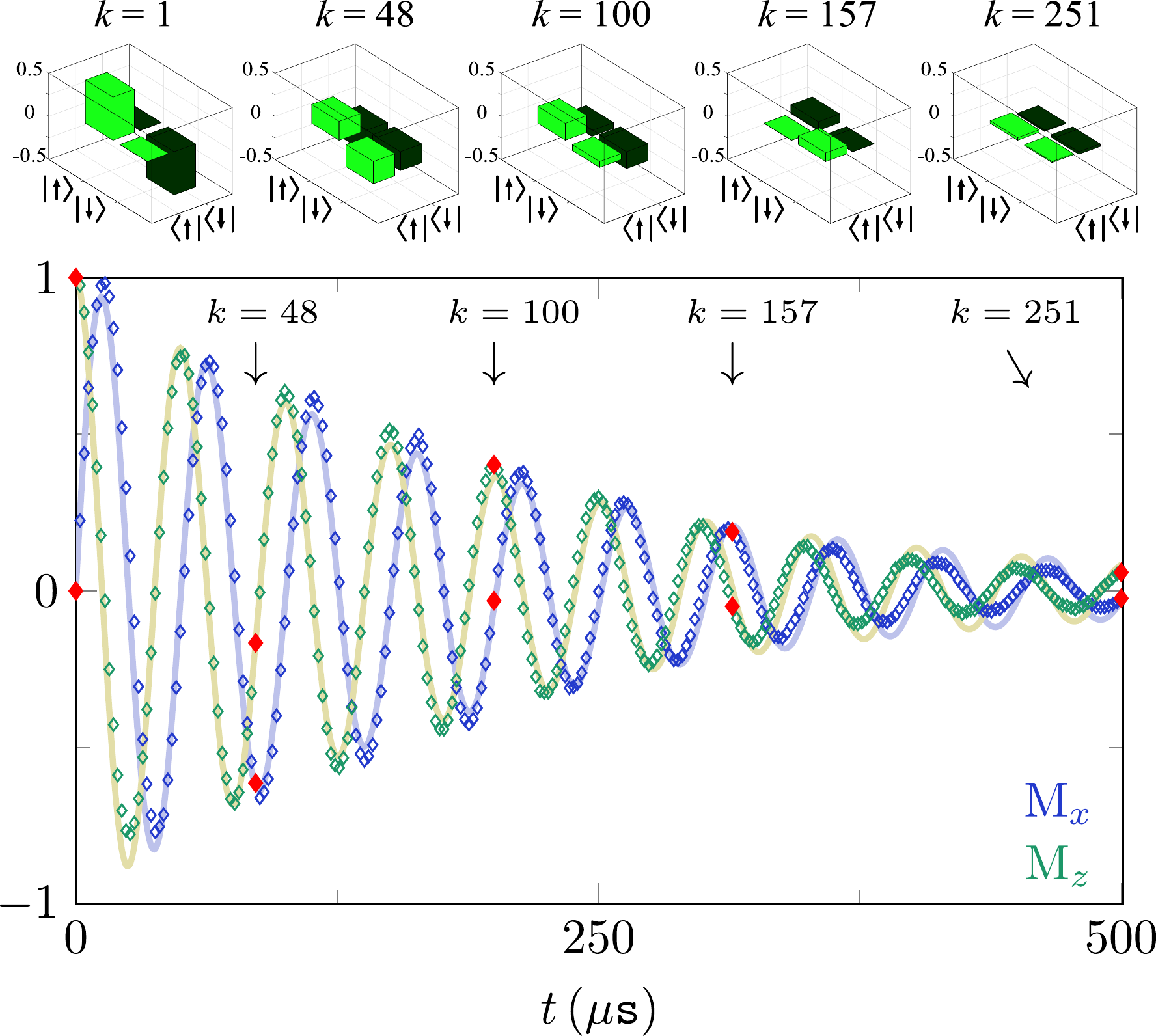} \\ [0.0cm]
     \mbox{ \textbf{(a)} } & \mbox{ \textbf{(b)} }
     \end{array}$
  \end{center}
	\caption{(Color online) On top, bar chart representing the real part of five experimental density matrices labelled by  $k=1,48,100,157,$ and 251 are depicted. On bottom, the experimental results (symbols) and theoretical prediction (solid lines) of the magnetization dynamics  are sketched. Numbered arrows denote density matrices and the respective values are represented by large red diamond symbols.  
(a) Data generated studying the TPP sample, and  performing fitting procedures the theoretical parameters were quantified   $\delta=11.5\,\mu$, $\mu / \omega_{1}=  3.95 \times 10^{-3}$, $\omega_{1}^{\text{th}}/\omega_{1}=1.05$ and $\nu=6.53\times10^{-2}$.
(b) Data generated studying the DSP sample, and  performing fitting procedures the theoretical parameters were quantified  $\delta=11.5\,\mu$, $\mu/\omega_{1}=  3.79 \times 10^{-3}$, $\omega_{1}^{\text{th}}/\omega_{1}=1.07$ and $\nu=5.82\times10^{-2}$.    } \label{fig:EXPMAGCOMP}
\end{figure*}

\begin{figure}[!ht]
  \begin{center}
    $\begin{array}{c}
     \multicolumn{1}{c}{\mbox{\bf }} \\ [-0.30cm]
     \epsfxsize=3.3000in
     \epsffile{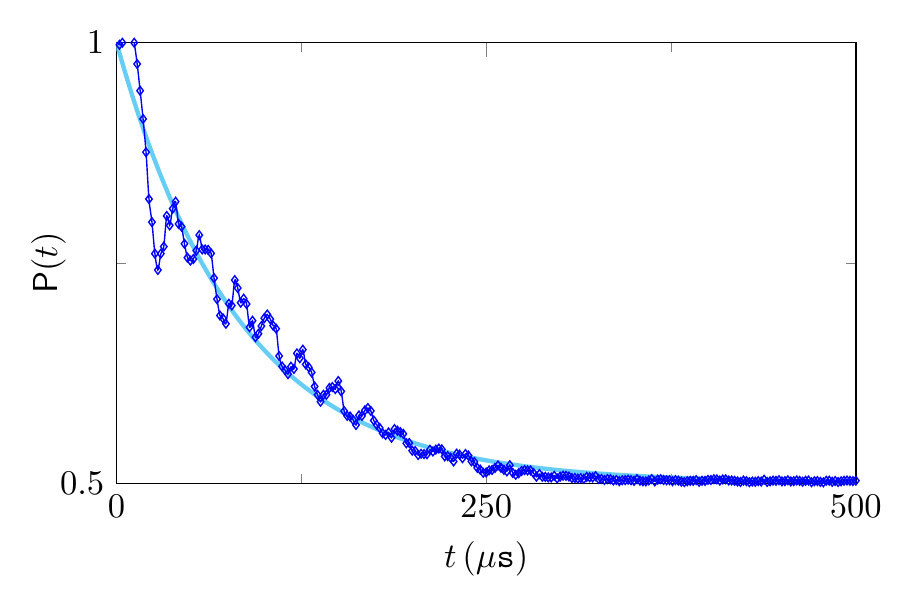} \\ [0.0cm]
     \mbox{ \textbf{(a)} } 
     \end{array}$
   $\begin{array}{c}
     \multicolumn{1}{c}{\mbox{\bf }} \\ [-0.30cm]
     \epsfxsize=3.3000in
     \epsffile{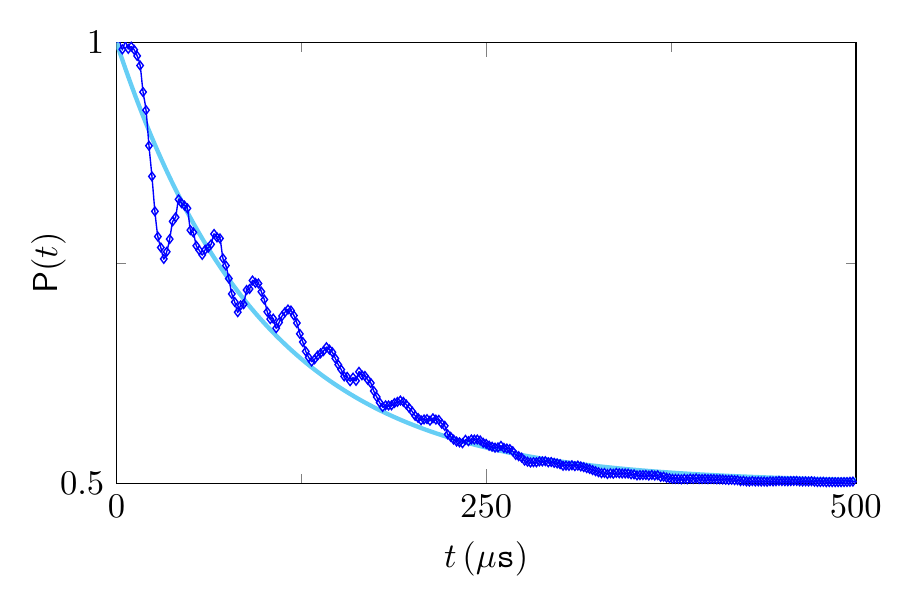} \\ [0.0cm]
     \mbox{ \textbf{(b)} } 
     \end{array}$
  \end{center}
	\caption{(Color online) Normalized purity $\mathsf{P}(t)$ in according to Eq. \eqref{purityf} for the dimensionless magnetization components in Eq. \eqref{Magnetizations}. It is plotted the purity calculated from the experimental data (symbols) and theoretical prediction (solid lines) for (a) TPP and (b) DSP.
    }    \label{fig:EXPPURITYCOMP}
\end{figure}

The nuclear magnetic moment operator $\hat{\mathbf{m}}$ is related to the angular momentum operator $\hat{\mathbf{I}}$ by means of the expression $\hat{\mathbf{m}} = \hbar\gamma \hat{\mathbf{I}}$. Thus, the mean value of the $k$-th component of the net magnetization is given by
\begin{equation}
\mathrm{m}_{k}(t) = \hbar\gamma\frac{\mathrm{\boldsymbol{Tr}}[\,\hat{\mathbf{I}}_{k}\hat{\varrho}(t)\,]}{\mathrm{\boldsymbol{Tr}}[\,\hat{\varrho}(t)\,]} , \label{Magnetization}
\end{equation}
with $k=x,y,z$.  At equilibrium, the net magnetization comes from the above definition as being $\mathrm{m}_{\text{eq}}=\hbar\gamma\mathrm{\mathbf{Tr}}[\, \hat{\mathbf{I}}_{z}\hat{\varrho}_{\text{eq}}\,] \approx \hbar^{2} \gamma \omega_{\text{L}} / 4k_{\text{B}}T$. Additionally, from the general formalism previously discussed in Sec. \ref{sec-unst},  Eq. (\ref{BlochVec}) allows defining the measurable dimensionless magnetization component $\mathrm{M}_{k}(t)$ for the pseudo-pure state evolving in time, in terms of the Bloch vector component $\mathrm{r}_{k}(t)$ as
\begin{equation}
\mathrm{M}_{k}(t)=\frac{\mathrm{m}_{k}(t)}{\mathrm{m}_{\text{eq}}} = \mathrm{r}_{k}(t). \label{DimensionlessMag}
\end{equation}
Therefore, these experimental details match  some theoretical definitions made in the previous section, such that the two most important are: first, we set the parameters of the Hamiltonian \eqref{H0} as    $\omega_{x}=\omega_{1}^{\text{th}}\cos{(\phi+\pi)}=0$, $\omega_{y}=\omega_{1}^{\text{th}}\sin{(\phi+\pi)}=\omega_{1}^{\text{th}}$, $\omega_{z}=-(\omega_{\text{L}}-\omega_{\text{rf}})=0$. Second, the dimensionless magnetization becomes 
\begin{subequations}
\label{Magnetizations}
\begin{align}
\mathrm{M}_{x}(t) & = [e^{-\delta t} + \nu\left(1-e^{-\mu t}\right)] \sin{\left(\omega_{1}^{\text{th}}\, t\right)}, \label{MagnetizationX} \\
\mathrm{M}_{z}(t) & = [e^{-\delta t} + \nu\left(1-e^{-\mu t}\right)]\cos{\left(\omega_{1}^{\text{th}}\, t\right)}, \label{MagnetizationZ}
\end{align}
\end{subequations}
whereas $\mathrm{M}_{y}(t)=0$. Here, $\nu$ means the resilient or residual magnetization of the sample, $\mu$ and $\delta$ are decay rates such that $\delta>\mu$. These parameters are given by fitting the experimental data in the theoretical model. The net magnetization components decrease over time as plotted in Fig.  \ref{fig:EXPMAGCOMP} as far as the system evolves from a pseudo-pure state to close maximally mixed state as in Eq. \eqref{purityf-longtimes} illustrated in Fig. \ref{fig:EXPPURITYCOMP}. 

The initial quantum state, represented by the density matrix  $ \hat{\varrho}(0)$, is graphically sketched using bar charts at the top of Fig. \ref{fig:EXPMAGCOMP}, labelled by $k=1$ and only the real part of  $ \hat{\varrho}(0)$ is plotted because the imaginary part assumes values close to null. The dynamics of the system is generated by the operator of Eq. (\ref{Utr}) such that the evolved density matrix at the time $t_{r}$ is denoted by  $\hat{\varrho} \left( t_{r} \right) $ where the tomography procedure was implemented at two hundred fifty one values of   $t_{r,k}$ $\in [ 0 \  \text{s},500\ \mu  \text{s} ]  $ with $\Delta t_{r}=t_{r,k}-t_{r,k-1}=2\ \mu  \text{s}$ and some density matrices with subscripts $k=48, \ 100, \ 157, \ 251$ are plotted and shown at the top of Fig. \ref{fig:EXPMAGCOMP}. The final quantum state is represented by the null density matrix operator, or at least the closest to it as experimentally possible.

\section{Discussion}  \label{sec-dis}

Formally, the definition of magnetization matches the  theoretical  definition of Bloch vector \cite{nielsen2000} or the pseudospin notation \cite{allen1975}. Therefore, it allows to transfer all previous theoretical description to the experimental setup  made for the magnetization of a one spin-half nuclear spin species \cite{bloch1946,oliveira2007,levitt2008}. In this sense, from the experimental  density matrices, $x,y,z$-magnetization components at each $t_{r,k}$ were computed and the data of the $x$- and $z$-components (symbols) are shown at bottom of Fig.  \ref{fig:EXPMAGCOMP}.  Similarly, the theoretical prediction (solid lines) of the $y$- and $z$-magnetization components are shown in Fig. \ref{fig:EXPMAGCOMP} and were computed using Eq. (\ref{Magnetizations}). Theoretical parameters $\delta$, $\mu$ and $\nu$ of both mathematical equations were evaluated performing fitting procedures from the TPP sample experimental data  $\delta=11.5\,\mu$,  $\mu/\omega_{1}=  3.95 \times 10^{-3}$ and $\nu=6.53\times10^{-2}$; and measuring the DSP sample  $\delta=11.5\,\mu$, $\mu/\omega_{1}=  3.79 \times 10^{-3}$ and $\nu=5.82\times10^{-2}$. 

The non-Hermitian approach introduced in this analysis considers a resilient magnetization $\nu$ at long time spin dynamics. This magnetization is characterised by the time regime $t\gg \delta^{-1} \approx 165.39 \,\mu\mathtt{s}$ for the TPP sample and $t\gg \delta^{-1}\approx 195.852 \,\mu\mathtt{s}$ for the  DSP sample.  This residual magnetization is due mainly to transversal magnetic field inhomogeneities along the sample. This signature happens similarly in solvent suppression NMR experiments in order to eliminate Hydrogen signals of water \cite{sklenar1987} or of   non-pure deuterated solvents  \cite{jesson1973,bax1985}. In this sense, the parameter value $\nu$ allows to quantify the percentage of this residual magnetization, which for both samples is approximately  $6.18 \pm 0.36 \,\%$  of the total magnetization, and apparently it is independent  of the sample preparation stoichiometry. To avoid this uncertainty, the residual magnetization of the TPP and DSP sample configures low and high concentration of $^{31}$P signal,  respectively. For both cases, the errors are corrected, on average, at the level of a few ($\sim 0.36$) percent. This percentage value could be improved, even closest to the null values, performing any of the suppression pulse technique \cite{jesson1973,bax1985,sklenar1987} or similar \cite{levitt2008}.

This theoretical non-Hermitian approach turns easy to handle with relaxation dynamics of an open quantum system, making it practical and favouring its direct application. On the other hand, the standard master equation approach is an efficient and a rigorous theoretical treatment of open quantum systems \cite{jones1966,jacquinot1973,smith1992,bull1992,palke1997}, but many times must be developed a laborious  and time-consuming mathematical effort to find the appropriate dynamical equations. Constrastingly, the non-Hermitian approach emerges as an alternative procedure to simplify any theoretical procedure to mimic the damped effect of the inhomogeneities of the time dependent radio-frequency as happens in the present discussion. This kind of theoretical discussion can be extended to a more recent NMR experimental setup  \cite{bengs2020}.

\begin{figure}[!ht]
  \begin{center}
    $\begin{array}{c}
     \multicolumn{1}{c}{\mbox{\bf }} \\ [-0.30cm]
     \epsfxsize=3.320in
     \epsffile{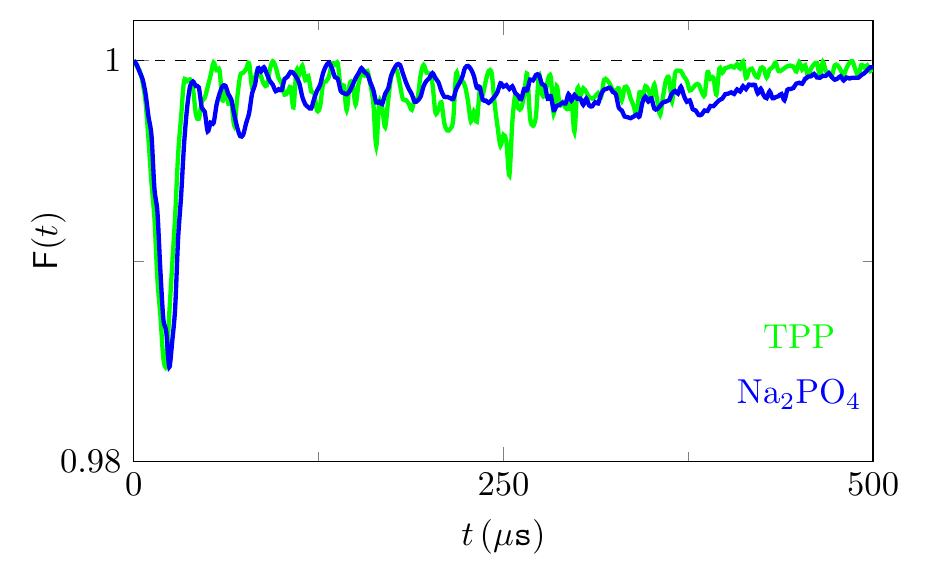} \\ [0.0cm]
     \end{array}$
  \end{center}
	\caption{(Color online) The fidelity parameter in according to Eqs. \eqref{Fidelity} is plotted for the TPP (green) and DSP (blue) samples. Both of them have a minimum close to $98,5\,\%$ in $22$ and $24\,\mu\mathtt{s}$ respectively for TPP and DSP. It increased slightly over the $30\,\mu \mathtt{s}$ and remains above $99,9 \,\%$, which means the theoretical model may precisely represent the experiment results.}  \label{fig:FIDELITY}
\end{figure}

The accuracy of the theoretical predictions may be checked by calculating the fidelity definition as discussed in \cite{fortunato2002}
\begin{equation}
\mathsf{F}(t_{r,k})=\frac{{\mathbf{Tr}}\left[ \hat{\varrho}_{\text{th}}(t_{r,k})\hat{\varrho}_{\text{exp}} (t_{r,k}) \right] }{\sqrt{ { \mathbf{Tr}} \left[ \hat{\varrho}_{\text{th}}^{2}(t_{r,k})\right] {\mathbf{Tr}} \left[ \hat{\varrho}_{\text{exp}}^{2}(t_{r,k}) \right] }},  
\label{Fidelity}
\end{equation}
where we compare the theoretical density operator $\hat{\varrho}_{\text{th}}(t_{r,k})$ to the experimental one $\hat{\varrho}_{\text{exp}}(t_{r,k})$. The time evolution of the fidelity parameter is shown in Fig. \ref{fig:FIDELITY}, in  which we obtain the minimum value of fidelity around $0.985$ (or 1.5 \% of error) at the time instant $22$ and $24\,\mu\mathtt{s}$ for TPP and DSP samples,  respectively. After decreasing slightly to the minimum value at the first time interval between 0 and $30\,\mu \mathtt{s}$, both systems remain the fidelity close to $1$, which means that the theoretical description matches the experimental data with a great accuracy.

\section{Concluding Remarks} \label{sec-conc}

The theoretical approach of non-Hermitian Hamiltonians is a theoretical framework that could be adapted to mimic some environment characteristics or modes of interaction with a quantum system. In this study, the non-Hermitian Hamiltonian  preserves the density operator properties, which implies probability conservation and normalization. Under the description of non-Hermitian Hamiltonian, the density matrix evolution  of $^{31}$P spin nuclei ensemble at any time could be useful for an experimental description of a long external radio-frequency pulse, or even in a continuous-wave irradiation regime. The residual magnetization at the end of the dynamics is a signature of the experimental implementation on solution NMR experiments, introduced in this study and denoted by the parameter $\nu$. Purity and fidelity definitions are used to warrant the spin ensemble dynamics' accuracy and the experimental implementations' quality, respectively. This analysis introduces the theory of non-Hermitian Hamiltonians as an alternative approach for relaxation processes that can be extended to other spin interactions like dipolar or quadrupolar ones by the NMR technique. 

\begin{acknowledgments}
The authors acknowledge the Brazilian agencies for financial support CAPES, CNPq Grant No. 453835/2014-7, 309023/2014-9 and 459134/2014-0; also  C-LABMU for kindly allowing the use of the NMR spectrometer. Part of the ideas were originated while D.C. was visiting the group of M.H.Y. Moussa at IFSC-USP, also D.C. is gratefully acknowledged for the generous hospitality. This study was financed in part by the Coordena\c{c}\~{a}o de Aperfei\c{c}oamento de Pessoal de Nível Superior - Brasil (CAPES) - Finance Code 001.
\end{acknowledgments}

\bibliographystyle{apsrev4-2}
\bibliography{references.bib}

\end{document}